\documentclass[11pt,a4paper]{article}
\usepackage[margin=2.5cm]{geometry}
\usepackage[T1]{fontenc}
\IfFileExists{lmodern.sty}{\usepackage{lmodern}}{}
\usepackage{amsmath,amssymb}
\usepackage{graphicx}
\usepackage{booktabs,tabularx,array}
\usepackage{xcolor}
\usepackage{dsfont} 
\usepackage[font=small,labelfont=bf]{caption}
\usepackage[colorlinks=true,linkcolor=blue!50!black,citecolor=blue!50!black,urlcolor=blue!50!black]{hyperref}

\newcolumntype{L}{>{\raggedright\arraybackslash}X}

\title{\textbf{Condensate: an interactive simulator of\\ two-dimensional quantum fluids}}
\author{Andrew W.\ Baggaley\\[2pt]
\small School of Mathematical Sciences, Lancaster University, Lancaster LA1 4YF, United Kingdom}
\date{}

\begin{document}
\maketitle

\begin{abstract}
\noindent
Condensate is an interactive simulator of two-dimensional quantum fluids that runs in a web browser on phones, tablets and computers, with no installation. It solves eight models built on the Gross--Pitaevskii equation: a single condensate, a dipolar gas with tilted dipoles, a binary mixture with quantum droplets, Rabi and spin--orbit coupled gases, a self-gravitating (Schr\"odinger--Poisson) condensate, a driven-dissipative polariton fluid, a rotating condensate and a finite-temperature stochastic projected Gross--Pitaevskii equation. A split-step Fourier solver runs on the graphics processor through WebGPU at up to $1024^2$ grid points, with a double-precision CPU engine as a fallback and reference. Users stir the fluid with a laser-like beam, imprint quantised vortices and change parameters while the fluid evolves. Each model comes with a ``Learn'' panel written for graduate students, giving the equations, core physics, and activities that pair an analytic prediction with an experiment in the app, and references keyed to a standard textbook. The whole application is one self-contained HTML file. In this accompanying text, we describe the models, the numerical method and the teaching material.
\end{abstract}

\section{Introduction}

Quantum fluids (and fluids in general) are best understood by watching them move, even better poking and prodding them to get a feel for how they respond. Indeed, phenomena such as vortex nucleation, the lattice of a rotating condensate, the unbinding of vortex pairs at the Berezinskii--Kosterlitz--Thouless (BKT) transition and the break-up of a dipolar gas into droplets are all dynamical and crucially can be captured in a two-dimensional model. Static figures convey this poorly, and the analysis that explains them is often beyond what students can do unaided. In a number of fields, interactive simulations have bridged a similar gap. However, typically research codes in the cold-atom community require installation, programming and offline visualisation, which puts them out of reach of new students.

This paper describes Condensate, an interactive simulator designed to let students produce and start to grasp these phenomena themselves, in real time, on a device they already carry. Its design follows a growing family of browser-based teaching tools: VisualPDE~\cite{visualpde} solves a broad class of user-defined partial differential equations on the GPU in real time; its authors argue that interactive ``play'' builds intuition that pen-and-paper analysis alone does not, especially for behaviour that is analytically intractable at undergraduate level. Earlier examples include the PhET simulations for physics teaching~\cite{phet}, Complexity Explorables~\cite{explorables}, the WebGL simulations of cardiac and nonlinear dynamics of Kaboudian, Cherry and Fenton~\cite{kaboudian} and, for quantum mechanics, the QuVis visualisations~\cite{quvis}. These tools share three features that Condensate adopts: no installation, immediate response to the user's input, and a curated set of models with sensible parameter ranges.

For quantum fluids, existing tools fall into two groups. Research codes such as GPELab~\cite{gpelab}, XMDS2~\cite{xmds}, GPUE~\cite{gpue} are powerful and well validated, but they are tools for research rather than teaching. Interactive tools exist but are narrower. The WebGL solver~\cite{stagg} of Stagg simulates the single-component two-dimensional GPE in the browser, and the super\_hydro explorer of Forbes and co-workers~\cite{superhydro} offers real-time interaction with superfluids through a Python and Jupyter environment that must be installed locally. Condensate sits between these groups. It provides:
\begin{itemize}
\item eight quantum-fluid models in one interface, sharing one numerical method and one set of units, so that students can compare, for example, a contact-interacting gas with a dipolar or self-gravitating one;
\item a WebGPU solver fast enough for a phone to run a $256^2$ grid interactively, packaged as one HTML file with no dependencies that also works offline;
\item interaction as part of the physics: the user's finger is a repulsive optical potential, and a tap imprints a quantised vortex;
\item teaching material for each model, with activities whose quantitative predictions have been checked against the simulation, and references keyed to the textbook of Barenghi, Bland and Parker~\cite{bbp}.
\end{itemize}

The intended users are lecturers and students in advanced undergraduate and postgraduate courses on superfluidity, cold atoms and nonlinear waves, and audiences for outreach. Researchers may also find it useful for building intuition about a model before writing a production code.

\section{Using Condensate}

\begin{figure}[t]
\centering
\includegraphics[height=6.3cm]{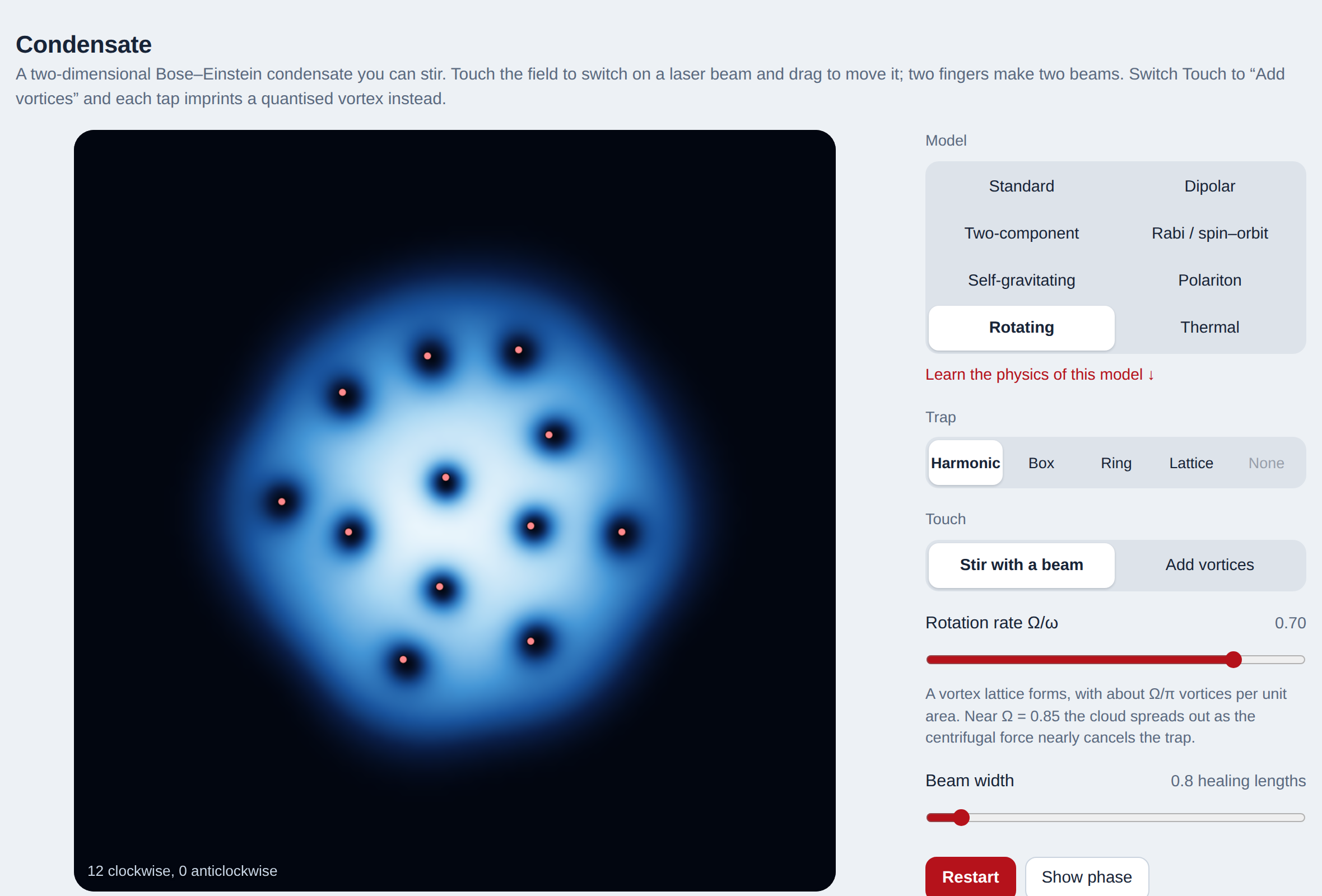}\hspace{1cm}
\includegraphics[height=6.3cm]{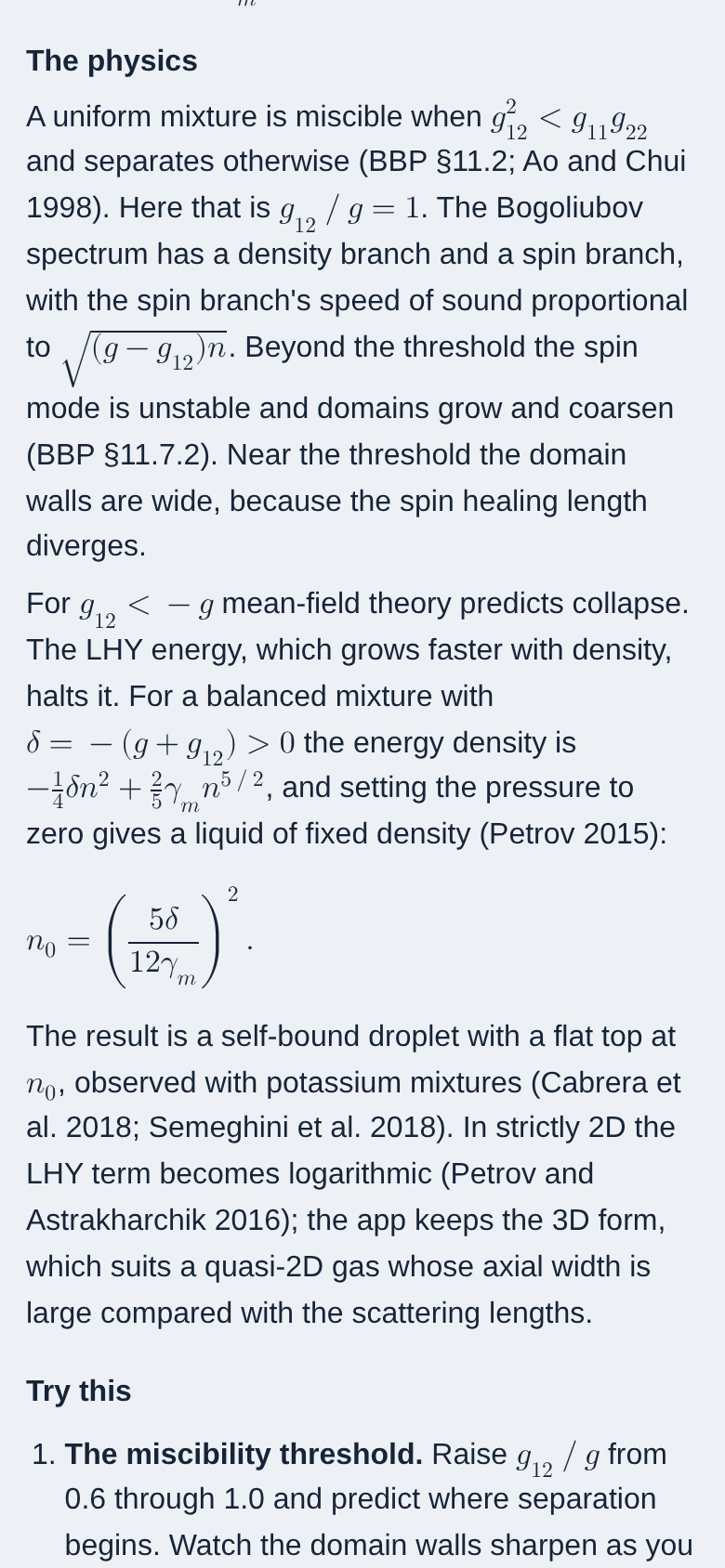}
\caption{Condensate on a computer (left) and a phone (right). Left: the rotating model at $\Omega=0.7\,\omega$, with vortices marked; the overlay reports 12 vortices of the same sign. Right: part of the Learn panel for the two-component model, with equations rendered natively as MathML.}
\label{fig:ui}
\end{figure}

Figure~\ref{fig:ui} shows the interface. The simulation fills a square panel. Brightness shows the density $n=|\psi|^2$ and an optional view shows the phase; for two-component models, the two species are shown in red and teal, and the phase view shows their relative phase. Touching the panel switches on a repulsive Gaussian beam at the touch point; dragging moves it, and two fingers make two beams. In ``Add vortices'' mode a tap instead imprints a quantised vortex of chosen sign. A ``Mark vortices'' button locates vortices from the phase winding around each grid cell and reports their number and signs.

The controls choose the model, the trap (harmonic with adjustable aspect ratio, box, ring, square optical lattice or none) and each model's main parameter, together with model-specific controls such as the dipole tilt, the spin--orbit coupling strength or the polariton noise and disorder. In most cases, parameters take effect immediately, so the fluid responds while it evolves. On browsers with WebGPU the grid size can be chosen from $128^2$ to $1024^2$; it defaults to $256^2$ on phones and $512^2$ on computers.

Below the simulation, a panel denoted ``Learn'' describes the current model (Section~\ref{sec:teaching}). It updates when the model changes, and its ``Set this up'' buttons configure the simulation for each suggested activity.

\section{Physical models}
\label{sec:models}

\subsection{Common framework}

\begin{figure}[t]
\centering
\includegraphics[width=\textwidth]{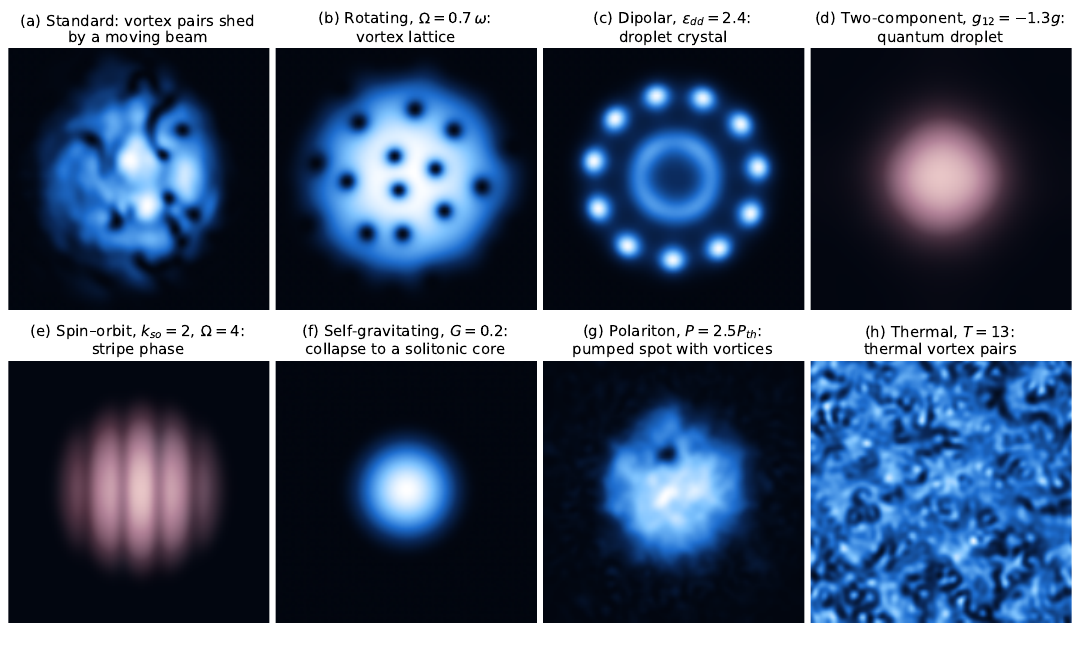}
\caption{Characteristic states of the eight models, from the app's CPU engine at $128^2$. In (d) and (e) the two species are shown in red and teal, as in the app; where they overlap the colours add. (g) and (h) show the time-averaged density displayed by the app.}
\label{fig:gallery}
\end{figure}

All models are written in trap units, $\hbar=m=\omega=1$, with lengths in $a_{\mathrm{ho}}=\sqrt{\hbar/m\omega}$ and times in $1/\omega$~\cite[\S3.8]{bbp}. The fluid is quasi-two-dimensional: the wavefunction is assumed frozen in a Gaussian profile of width $l_z=1.5$ along $z$, which sets the 2D contact coupling $g$ and the averaging of the long-range kernels below. Each model evolves a generalised, optionally damped, GPE,
\begin{equation}
(i-\gamma)\,\partial_t\psi = \Bigl[-\tfrac12\nabla^2 + V(\mathbf r) + V_{\mathrm{beam}}(\mathbf r,t) + g|\psi|^2 + \Phi[\psi] - \mu\Bigr]\psi,
\label{eq:gpe}
\end{equation}
where $\gamma\ge0$ is a phenomenological damping~\cite{choi}, $V_{\mathrm{beam}}=V_0\,e^{-|\mathbf r-\mathbf r_b(t)|^2/2\sigma^2}$ is the user's beam, with $V_0=60$ well above the chemical potential, and $\Phi[\psi]$ collects the model-specific terms. At the default $128^2$ grid, $g=1$ and $N=475$, so the Thomas--Fermi chemical potential in the harmonic trap is $\mu=\sqrt{gN/\pi}\approx12.3$, the healing length is $\xi=1/\sqrt{gn_0}\approx0.28$ and the central sound speed is $c\approx3.6$. Table~\ref{tab:models} summarises the eight models, and Fig.~\ref{fig:gallery} shows a characteristic state of each.

\begin{table}[t]
\centering\small
\begin{tabularx}{\textwidth}{@{}p{2.3cm}LLL@{}}
\toprule
Model & Additional physics & Controls & Phenomena \\
\midrule
Standard & none & damping $\gamma$ & critical velocity, vortex shedding, vortex dipoles, sound \\
Rotating & $-\Omega L_z$ (rotating frame) & $\Omega/\omega\in[0,0.85]$ & vortex entry, Abrikosov lattice, Feynman's rule \\
Dipolar & quasi-2D dipolar kernel with tilt $\alpha$; Lee--Huang--Yang term & $\varepsilon_{dd}\in[0,2.4]$, tilt $\alpha$, azimuth $\phi$ & roton, droplet crystals, magnetostriction, magic angle \\
Two-component & cross-interaction $g_{12}$; mixture Lee--Huang--Yang term & $g_{12}/g\in[-1.5,1.6]$, species share & miscibility, domains, filled cores, quantum droplets \\
Rabi / spin--orbit & $\tfrac12(p_x-k_{so}\sigma_z)^2+\tfrac12p_y^2+\tfrac{\Omega}{2}\sigma_x$ & $\Omega$, $k_{so}$, $g_{12}/g$ & stripe, plane-wave and zero-momentum phases; confined vortices \\
Self-gravitating & Schr\"odinger--Poisson, quasi-2D & $G\in[0,0.4]$ & Jeans instability, solitonic cores \\
Polariton & driven-dissipative GPE with reservoir; noise; disorder & $P/P_{th}$, noise, disorder & threshold, outflow, spontaneous and pinned vortices \\
Thermal & simple-growth SPGPE at fixed $\mu$ & temperature $T$ & thermal vortex pairs, BKT physics \\
\bottomrule
\end{tabularx}
\caption{The eight models. Each solves Eq.~\eqref{eq:gpe} with the additional terms listed.}
\label{tab:models}
\end{table}

\subsection{Model-specific terms}

\paragraph{Dipolar.} For dipoles tilted by $\alpha$ from the $z$ axis, with their in-plane projection along $\hat{\mathbf e}_\parallel=(\cos\phi,\sin\phi)$, averaging the dipole--dipole interaction over the axial profile gives the Fourier-space kernel~\cite{fischer,ticknor}
\begin{equation}
\tilde U(\mathbf k) = g + g\varepsilon_{dd}\Bigl[\cos^2\alpha\,\bigl(2-3F\bigr) + \sin^2\alpha\Bigl(\frac{3Fk_\parallel^2}{k^2}-1\Bigr)\Bigr],\qquad k_\parallel=k_x\cos\phi+k_y\sin\phi,
\end{equation}
with $F=\sqrt\pi\,q\,e^{q^2}\operatorname{erfc}q$ and $q=kl_z/\sqrt2$. At long wavelengths this reduces to $g[1+\varepsilon_{dd}(3\cos^2\alpha-1)]$, so the isotropic dipolar contribution vanishes at the magic angle $\alpha_m=\arccos(1/\sqrt3)\approx54.7^\circ$~\cite{giovanazzi}. A Lee--Huang--Yang (LHY) term $\gamma_{\mathrm{LHY}}(1+\tfrac32\varepsilon_{dd}^2)|\psi|^3$ with $\gamma_{\mathrm{LHY}}=0.02$, of the three-dimensional form~\cite{limapelster}, prevents collapse and stabilises droplets~\cite{kadau,chomaz}.

\paragraph{Two-component.} Two species with equal self-interaction $g$ and cross-interaction $g_{12}$ mix for $g_{12}<g$ and separate for $g_{12}>g$~\cite{aochui}. Each species also feels $\gamma_m(|\psi_1|^2+|\psi_2|^2)^{3/2}$ with $\gamma_m=0.04$. For $\delta=-(g+g_{12})>0$ the balanced mixture has energy density $-\tfrac14\delta n^2+\tfrac25\gamma_m n^{5/2}$, and zero pressure gives a self-bound liquid of density $n_0=(5\delta/12\gamma_m)^2$, Petrov's quantum droplet~\cite{petrov,cabrera,semeghini}. In strictly two dimensions the LHY term becomes logarithmic~\cite{petrovastrakharchik}; the three-dimensional form used here describes a quasi-2D gas whose axial width is large compared with the scattering lengths.

\paragraph{Rabi and spin--orbit coupling.} The single-particle Hamiltonian is that of the NIST experiments~\cite{lin}, whose lower band has minima at $\pm k_1=\pm k_{so}\sqrt{1-(\Omega/2k_{so}^2)^2}$ for $\Omega<2k_{so}^2$. Interactions select the stripe phase ($g_{12}<g$), the plane-wave phase ($g_{12}>g$) or, above $\Omega=2k_{so}^2$, the zero-momentum phase~\cite{hozhang,lipitaevskiistringari}. The spin--orbit term only shifts each species' kinetic energy to $\tfrac12k^2\mp k_{so}k_x$, and the Raman coupling is an exact local rotation, so both remain exact in the split-step scheme. With $k_{so}=0$ the model is a coherently coupled mixture, in which a vortex in one species drags a domain wall of relative phase~\cite{sonstephanov}.

\paragraph{Self-gravitating.} Newtonian gravity averaged over the axial profile gives
\begin{equation}
\tilde\Phi(\mathbf k)=-G\,\frac{2\pi}{k}\operatorname{erfcx}\Bigl(\frac{kl_z}{\sqrt2}\Bigr)\,\tilde n(\mathbf k),
\end{equation}
with the uniform background removed as in cosmological codes. This is the Schr\"odinger--Poisson system used for ultralight dark matter~\cite{hui2017,schive}, here with an optional contact term.

\paragraph{Polariton.} The driven-dissipative GPE of Wouters and Carusotto~\cite{woucar} couples the condensate to an incoherent reservoir,
\begin{equation}
i\partial_t\psi = \Bigl[-\tfrac12\nabla^2+V+g|\psi|^2+g_Rn_R+\tfrac i2(Rn_R-\gamma_c)\Bigr]\psi,\qquad \partial_tn_R = P-(\gamma_R+R|\psi|^2)\,n_R ,
\end{equation}
with threshold $P_{th}=\gamma_c\gamma_R/R$ and, for uniform pumping, steady density $n=(P-P_{th})/\gamma_c$~\cite{carusottociuti}. The trap buttons shape the pump. Optional truncated-Wigner noise~\cite{wousavona} and a Gaussian-correlated disorder potential model a real microcavity.

\paragraph{Rotating.} The contact GPE is solved in the frame rotating at $\Omega$, with $-\Omega L_z$ added to the Hamiltonian; with no trap the rotating frame is unbounded, so the trap is required.

\paragraph{Thermal.} The simple-growth stochastic projected GPE~\cite{gardinerdavis,bradley,blakie},
\begin{equation}
d\psi=\mathcal P\bigl\{-(i+\gamma)(L-\mu)\psi\,dt+dW\bigr\},\qquad \langle dW^*dW\rangle=2\gamma T\,dt,
\end{equation}
with $L=-\tfrac12\nabla^2+V+g|\psi|^2$, couples the low-energy modes $|\mathbf k|<k_c=6.5$ to a reservoir at temperature $T$ and chemical potential $\mu=12$. As in all classical-field methods, the temperature at which vortex pairs unbind depends on the cutoff.

\subsection{Scaling with grid size}

The grid spacing is fixed at $17/128$ in trap units, so a larger grid simulates a larger system with the same local physics rather than a finer one. On a grid $s=N/128$ times wider, the harmonic frequency becomes $1/s$, the box and ring radii and the polariton spot grow by $s$, and the atom number grows as $s^2$, which keeps the density, chemical potential and healing length fixed. The harmonic trap can be made anisotropic, $V=\tfrac12(\omega_x^2x^2+\omega_y^2y^2)$ with $\lambda=\omega_y/\omega_x\in[1,6]$; since the two-dimensional Thomas--Fermi atom number is $N=\pi\mu^2/(g\omega_x\omega_y)$, the atom number is divided by $\lambda$, which again keeps $\mu$ and $\xi$ fixed and produces cigar-shaped clouds of fixed length. Local features keep their physical size: the box wall, the lattice period, the disorder correlation length and the ring's cross-section. The ring therefore grows only in circumference, and its atom number scales as $s$. Rotation and gravity are scaled by $1/s$ for the same reason.

\subsection{Trapping potentials}
\label{sec:traps}

The trap buttons choose the external potential $V(\mathbf r)$ in Eq.~\eqref{eq:gpe}, written here for a grid $s=N/128$ times the default width, with $r=|\mathbf r|$. In every case the potential is capped at $V_{\max}=100$. The cap only affects regions far outside the fluid, where the density is negligible, and it bounds the phase that $V$ adds in a single time step.

\paragraph{Harmonic.} A harmonic trap, optionally anisotropic,
\begin{equation}
V_{\mathrm{h}}(\mathbf r)=\tfrac12\bigl(\omega_x^2x^2+\omega_y^2y^2\bigr),\qquad \omega_x=\frac{1}{s},\quad \omega_y=\frac{\lambda}{s},
\end{equation}
where the aspect slider sets $\lambda=\omega_y/\omega_x\in[1,6]$. At $\lambda=1$ this is the isotropic trap that defines the units, $\tfrac12r^2$ on the default grid. The atom number is divided by $\lambda$ (Section~\ref{sec:models}), so that the Thomas--Fermi radii $R_i=\sqrt{2\mu}/\omega_i$ change while $\mu$ does not.

\paragraph{Box.} A flat-bottomed disc with a smooth wall,
\begin{equation}
V_{\mathrm{b}}(\mathbf r)=\frac{V_{\mathrm{w}}}{2}\Bigl[1+\tanh\Bigl(\frac{r-R_{\mathrm b}}{w}\Bigr)\Bigr],\qquad V_{\mathrm w}=40,\quad R_{\mathrm b}=5.5\,s,\quad w=0.25,
\end{equation}
modelled on the uniform optical box traps of recent experiments~\cite{gaunt}. In the box the density is $n\approx N/\pi R_{\mathrm b}^2$, so $\mu\approx5$ on the default grid; the wall is about eight times higher than $\mu$ and narrower than the local healing length, $\xi\approx0.45$, so it is effectively hard, while the $\tanh$ profile avoids the ringing that a step would produce in a spectral method.

\paragraph{Ring.} An annular trap, harmonic in the radial direction,
\begin{equation}
V_{\mathrm{r}}(\mathbf r)=\tfrac12\,\omega_r^2\,(r-R_{\mathrm r})^2,\qquad \omega_r=3,\quad R_{\mathrm r}=4.5\,s,
\end{equation}
the geometry of persistent-current experiments~\cite{ramanathan}. The radial frequency, and so the ring's cross-section, is fixed, so only the circumference grows on larger grids, and the atom number grows as $s$ rather than $s^2$.

\paragraph{Lattice.} A square optical lattice on a harmonic envelope,
\begin{equation}
V_{\mathrm{L}}(\mathbf r)=\frac{r^2}{2s^2}+V_0^{\mathrm L}E_R\Bigl[\sin^2\Bigl(\frac{\pi x}{a}\Bigr)+\sin^2\Bigl(\frac{\pi y}{a}\Bigr)\Bigr],\qquad E_R=\frac12\Bigl(\frac{\pi}{a}\Bigr)^2,\quad a=1,
\end{equation}
with lattice period $a$, the recoil energy $E_R=\hbar^2k_L^2/2m$ of the lattice wavenumber $k_L=\pi/a$, and a depth $V_0^{\mathrm L}\in[0,10]$ in recoil energies set by a slider~\cite[\S3.9.3]{bbp}, \cite{morsch}. On the default grid the period equals the oscillator length, and on every grid an integer number of periods ($17s$) fits the box, so the lattice is consistent with the periodic boundary conditions. With $E_R=\pi^2/2\approx4.9$, the default depth of $3E_R$ is comparable to $\mu$, between the nearly free and tight-binding regimes.

\paragraph{None.} $V=0$. The periodic box then approximates an infinite uniform gas, which is the natural setting for droplets, the Jeans instability and thermal vortex pairs. The rotating model requires a trap, since the rotating frame is unbounded without one.

\paragraph{Polariton pump profiles.} In the polariton model the same buttons also shape the pump, $P(\mathbf r)=p\,P_{th}\,f(\mathbf r)$, where $p$ is the pump slider and
\begin{equation}
f(\mathbf r)=\begin{cases}
e^{-r^2/2\sigma_p^2},\quad \sigma_p=3.5\,s & \text{harmonic and lattice (Gaussian spot)},\\[2pt]
\tfrac12\bigl[1-\tanh\bigl((r-R_{\mathrm b}+0.4)/0.3\bigr)\bigr] & \text{box (top hat)},\\[2pt]
e^{-(r-R_{\mathrm r})^2/2w_p^2},\quad w_p=0.9 & \text{ring (annulus)},\\[2pt]
1 & \text{none (uniform)}.
\end{cases}
\end{equation}
The condensate forms where $P(\mathbf r)>P_{th}$ and, in the local-density approximation, has density $n(\mathbf r)\approx(\gamma_R/R)\,[P(\mathbf r)/P_{th}-1]$, so the pumped region appears as the bright part of the fluid, and the condensate ends roughly where the pump falls below threshold. The trap potential still acts alongside the pump, and the disorder potential (Section~\ref{sec:numerics}) is added to $V$ when its slider is non-zero.

\section{Numerical method}
\label{sec:numerics}

\subsection{Discretisation}

The wavefunction lives on an $N\times N$ periodic grid covering a square of side $L=17s$, with $s=N/128$, so the grid spacing is fixed at $\Delta x=17/128\approx0.133$ (Section~\ref{sec:models}). Spatial derivatives are spectral: the discrete Fourier transform $\mathcal F$ represents $\psi$ by the wavenumbers $k_{x,y}=2\pi m/L$ with $-N/2\le m<N/2$, up to the Nyquist wavenumber $k_{\max}=\pi/\Delta x\approx23.7$. The healing length at the default density, $\xi\approx0.28$, spans about two grid spacings, and a vortex core is resolved by several points. The wavefunction is normalised to the atom number, $\sum|\psi_j|^2\Delta x^2=N_{\mathrm{at}}$.

\subsection{Split-step integrator}

Every model is advanced by Strang splitting~\cite{strang,bao} of the generator into a local part, diagonal in real space, and a kinetic part, diagonal in Fourier space. Specifically, we integrate
\begin{equation}
\partial_t\psi=-(i+\gamma)\,(H-\mu)\,\psi,
\label{eq:damped}
\end{equation}
which is Eq.~\eqref{eq:gpe} with time rescaled by $1+\gamma^2$, a negligible difference for $\gamma\le0.1$. With $H=T+W$, where $T=-\tfrac12\nabla^2$ and $W$ collects every local term, one step of length $\Delta t$ is
\begin{equation}
\psi(t+\Delta t)=\mathcal L\bigl(\tfrac12\Delta t\bigr)\,\mathcal K(\Delta t)\,\mathcal L\bigl(\tfrac12\Delta t\bigr)\,\psi(t),
\end{equation}
with
\begin{align}
\mathcal L(h)\,\psi &= \exp\bigl[-(i+\gamma)\,\bigl(W(\mathbf r)-\mu\bigr)\,h\bigr]\,\psi, \\
\mathcal K(\Delta t)\,\psi &= \mathcal F^{-1}\Bigl[f(k)\,\exp\bigl(-(i+\gamma)\,\tfrac12k^2\Delta t\bigr)\,\mathcal F\psi\Bigr].
\end{align}
The local potential is
\begin{equation}
W = V + V_{\mathrm{beam}} + g|\psi|^2 + \Phi + W_{\mathrm{LHY}} + W_{\mathrm{model}},
\end{equation}
where $W_{\mathrm{LHY}}$ is the relevant Lee--Huang--Yang term and $W_{\mathrm{model}}$ holds the cross-interaction, reservoir or disorder terms of the multi-component and polariton models. Each half-step recomputes $|\psi|^2$ before applying $\mathcal L$, so the contact nonlinearity is treated to second order in $\Delta t$; for $\gamma=0$ it is exact, because a local phase rotation leaves $|\psi|^2$ unchanged. Each factor is unitary when $\gamma=0$, so the scheme conserves the norm and has no stability limit from the kinetic or contact terms. The real-time step is $\Delta t=0.008$, except for tilted dipoles (Section~\ref{sec:tilt}).

\paragraph{Chemical potential and normalisation.} In closed systems the wavefunction is rescaled after every step to restore the atom number, which makes Eq.~\eqref{eq:damped} the usual damped GPE at fixed $N_{\mathrm{at}}$. The chemical potential is estimated as the density-weighted mean of the local potential, $\mu\approx\sum_j n_jW_j/\sum_jn_j$, which keeps the phase rotation of the condensate slow; it is updated every tenth step on the CPU and every step on the GPU, from the previous step's reductions. Because the norm is restored anyway, $\mu$ only sets the reference energy for the damping.

\paragraph{Spectral filter.} The kinetic factor includes the filter $f(k)=\exp[-1.5\,(k/k_{\max})^{12}]$. It is above $0.99$ for $k<0.66\,k_{\max}$, falls to $0.22$ at $k_{\max}$, and removes the corners of the Fourier square, where $|\mathbf k|$ reaches $\sqrt2\,k_{\max}$. Applied once per step, it acts as a weak hyperviscosity that controls the build-up of aliased power from the nonlinear terms, which generate wavenumbers beyond those the grid can represent.

\subsection{Nonlocal interactions}

The dipolar and gravitational potentials are convolutions of the density with a kernel, and are evaluated with the convolution theorem,
\begin{equation}
\Phi(\mathbf r)=\mathcal F^{-1}\bigl[\tilde U(\mathbf k)\,\mathcal F[n](\mathbf k)\bigr],\qquad n=|\psi|^2,
\end{equation}
at a cost of two extra two-dimensional FFTs per step. The kernels $\tilde U(\mathbf k)$ of Section~\ref{sec:models} are tabulated on the grid when a parameter changes. The function $\operatorname{erfcx}q=e^{q^2}\operatorname{erfc}q$ in both kernels is evaluated with a rational approximation for $q<2$ and a continued fraction for $q\ge2$. For the dipolar kernel the $\mathbf k=0$ mode, $g\varepsilon_{dd}(3\cos^2\alpha-1)$, is retained (the mean-field shift of a uniform dipolar gas); for self-gravity the $\mathbf k=0$ mode is set to zero, which removes the interaction with the uniform background, as is standard in cosmological codes.

The transform is periodic, so each cloud also interacts with its periodic images. The dipolar interaction falls off as $1/r^3$ and the clouds fill only part of the box, so the image interaction is small; for gravity the leading error is a tidal term of relative order $(R/L)^3$ for a cloud of radius $R$. Whilst zero padding would remove these errors (at four times the computational cost), we judged it unnecessary for a teaching tool.

$\Phi$ is computed once per step, from the density at the start of the step, and held fixed through both half-steps. This saves two FFTs per step but makes the nonlocal coupling first order in time.

\subsection{Time step for tilted dipoles}
\label{sec:tilt}

Tilting the dipoles makes the kernel strongly repulsive at short wavelengths along their in-plane direction, where $\tilde U\to g+2g\varepsilon_{dd}$, whereas at $\alpha=0$ it tends to $g-g\varepsilon_{dd}$. Because $\Phi$ is treated explicitly, the scheme is then only conditionally stable, so the dipolar step is reduced adaptively to
\begin{equation}
\Delta t=\frac{0.008}{1+0.8\,\varepsilon_{dd}\sin^2\alpha},
\end{equation}
which leaves $\alpha=0$ unchanged and reaches $0.0027$ at $\varepsilon_{dd}=2.4$, $\alpha=90^\circ$. The same step is stable for every azimuth $\phi$, including the grid diagonal, where the largest wavenumbers are $\sqrt2$ times those along an axis. Strongly tilted, strongly dipolar runs therefore advance up to three times more slowly.

\subsection{Rotation}

In the rotating frame the Hamiltonian gains $-\Omega L_z=\Omega\,y\,p_x-\Omega\,x\,p_y$. Combined with the kinetic energy, this term is diagonal in neither real nor Fourier space, so the kinetic step is replaced by an alternating-direction split~\cite{baoduzhang},
\begin{equation}
T-\Omega L_z = \underbrace{\tfrac12p_x^2+\Omega\,y\,p_x}_{A} + \underbrace{\tfrac12p_y^2-\Omega\,x\,p_y}_{B},
\end{equation}
in which $A$ is diagonal in $(k_x,y)$ and $B$ in $(x,k_y)$. The rotating kinetic step is then
\begin{equation}
\mathcal K_\Omega(\Delta t)=e^{-(i+\gamma)A\Delta t/2}\,e^{-(i+\gamma)B\Delta t}\,e^{-(i+\gamma)A\Delta t/2},
\end{equation}
nested inside the outer Strang split. Each factor is applied with one-dimensional transforms: $e^{-(i+\gamma)Ah}$ by a forward FFT along $x$ for every row of fixed $y$, multiplication by $\exp[-(i+\gamma)(\tfrac12k_x^2+\Omega yk_x)h]$ and an inverse FFT along $x$; $B$ likewise along $y$ for every column of fixed $x$. The three factors cost six sets of one-dimensional transforms, the equivalent of one and a half two-dimensional FFT pairs, and each is unitary when $\gamma=0$, so the scheme remains norm-preserving and unconditionally stable. The spectral filter is applied in each direction separately. Because $[A,B]\neq0$ the inner split has its own second-order error, of the same order as the outer one.

\subsection{Two-component and coupled models}

In the two-component model each species has its own kinetic step and local potential, the latter including $g_{12}|\psi_{3-j}|^2$ and the mixture LHY term $\gamma_m(n_1+n_2)^{3/2}$; both local potentials are evaluated from the densities before either species is updated. Each species is normalised separately.

In the coupled model the spin--orbit term is diagonal in Fourier space and only shifts the kinetic energy of the two species to $T_\pm=\tfrac12k^2\mp k_{so}k_x$, which the kinetic factor includes exactly. The Raman coupling $\tfrac12\Omega\sigma_x$ is a local $2\times2$ operator whose exponential is known in closed form,
\begin{equation}
e^{-(i+\gamma)\frac{\Omega}{2}h\,\sigma_x}=\cosh\bigl[(i+\gamma)\tfrac{\Omega}{2}h\bigr]\,\mathds{1}-\sinh\bigl[(i+\gamma)\tfrac{\Omega}{2}h\bigr]\,\sigma_x ,
\end{equation}
which reduces to a rotation, $\cos(\tfrac12\Omega h)\mathds{1}-i\sin(\tfrac12\Omega h)\sigma_x$, when $\gamma=0$. It is applied after the local factor in the first half-step and before it in the second, so the full step remains second order; the coupling moves atoms between species, only the total atom number is restored, and the two species share one chemical potential.

\subsection{Ground states}

Initial states are found by imaginary-time evolution~\cite{chiofalo}: the same split-step factors with $i+\gamma$ replaced by $1$ and $\mu=0$, and the norm restored after every step, a normalised gradient flow. The imaginary step is $\Delta\tau=0.004$ ($0.01$ for the coupled model), for 500 steps (900 rotating, 600 coupled). The initial guess is the Thomas--Fermi profile $n=\max(0,\mu-V)/g_{\mathrm{eff}}$, with $\mu$ found by bisection and an effective coupling that includes the long-wavelength interaction: $g[1+\varepsilon_{dd}(3\cos^2\alpha-1)]$ for dipoles and $\tfrac12g(1+g_{12}/g)$ for balanced mixtures. A random perturbation of up to $\pm3\%$ breaks symmetries. Three models need a more specific seed. The rotating model imprints the Feynman number of vortices, $\Omega R^2$ for a cloud of radius $R$, at random positions, so that cooling relaxes them into a lattice instead of waiting for them to enter through the surface. The coupled model starts from lower-band spinors at $\pm k_1$ with unequal weights, so that cooling can select a stripe or plane-wave phase. A self-bound droplet starts as a flat-topped disc at the predicted density $n_0$. After cooling, a small amount of noise is added so that real-time instabilities can develop.

\subsection{Open and stochastic models}

\paragraph{Polariton.} The gain and loss enter the local factor as the real exponential $\exp[\tfrac12(Rn_R-\gamma_c)h]$, alongside the potential. The reservoir obeys a linear equation once $|\psi|^2$ is frozen over a step, and is advanced exactly,
\begin{equation}
n_R(t+\Delta t)=n_{\mathrm{eq}}+\bigl(n_R(t)-n_{\mathrm{eq}}\bigr)e^{-\kappa\Delta t},\qquad \kappa=\gamma_R+R|\psi|^2,\quad n_{\mathrm{eq}}=P/\kappa ,
\end{equation}
an exponential integrator that is stable for any step. The kinetic factor carries a weak damping, $\gamma=0.01$, which models energy relaxation. The wavefunction is not renormalised: its amplitude is set by the balance of gain and loss. The condensate grows from a smooth random seed, built from modes with $|\mathbf k|<3$, so that it forms in real time.

\paragraph{Stochastic terms.} The thermal model and the optional polariton noise add complex Gaussian white noise once per step, in Fourier space after the kinetic factor, to the modes with $|\mathbf k|<k_c=6.5$. The noise has real-space variance $2D\Delta t/\Delta x^2$ per grid point, with $D=\gamma T$ for the thermal model, matching $\langle dW^*dW\rangle=2\gamma T\,dt$, and $D=\sigma\gamma_c/2$ for polaritons. This is an Euler--Maruyama treatment of the noise within the split step. In the thermal model the projector $\mathcal P$ replaces the spectral filter: the kinetic factor is multiplied by the indicator of $|\mathbf k|<k_c$, so the noise and the nonlinear term never populate modes above the cutoff. The damping $\gamma=0.05$ appears in both the local and kinetic factors, through $i+\gamma$, and $\mu$ is fixed at $12$, so the atom number is set by the reservoir rather than imposed. Both noisy models display a running average of the density, $\bar n\leftarrow\bar n+(\Delta t/\tau)(|\psi|^2-\bar n)$ with $\tau=0.25$, which plays the role of a camera's exposure time.

\paragraph{Disorder.} The polariton disorder potential is a Gaussian random field with correlation
\begin{equation}
\langle V(0)V(\mathbf r)\rangle\propto e^{-r^2/\ell^2},\qquad \ell=0.6.
\end{equation} It is drawn in Fourier space with amplitude $e^{-k^2\ell^2/8}$ and random phase, transformed back, and rescaled to zero mean and unit variance; the slider sets its rms amplitude.

\subsection{Interaction}

The beam is the potential $V_{\mathrm{beam}}=V_0\sum_b a_b(t)\exp[-|\mathbf r-\mathbf r_b(t)|^2/2\sigma^2]$, which enters $W$. Each beam centre $\mathbf r_b$ follows the touch point at a speed of at most 4 trap units, which is faster than the central sound speed, and its amplitude $a_b$ ramps between 0 and 1 over 0.2 time units, so that switching a beam on or off does not deliver an impulsive kick. A tap in vortex mode multiplies the wavefunction by $r/\sqrt{r^2+2\xi^2}\,e^{\pm i\theta}$, centred on the tap, with $\xi$ from the local density, which imprints a vortex with approximately the right core profile. Vortices are located by summing the phase differences around each grid plaquette and keeping windings of $\pm2\pi$ where the smoothed density exceeds a fifth of its maximum, which rejects phase defects in the low-density edge.

\subsection{Implementation}

\paragraph{GPU engine.} The GPU engine implements the whole loop in WGSL compute shaders, in single precision with complex numbers stored as pairs of 32-bit floats. We note that the built-in WGSL sine and cosine are only guaranteed to about $5\times10^{-4}$. Hence, the GPU engine uses its own hand-coded sine and cosine, with argument reduction by $\pi/2$ in two parts followed by minimax polynomials, accurate to about $10^{-7}$. This ensured good agreement with the CPU approach.

\paragraph{CPU engine.} The CPU engine performs the same operations in double precision at $128^2$, with an in-place radix-2 FFT applied along rows and, after a transpose, along columns. It caches the kinetic and rotation factors and recomputes them only when a parameter changes. It runs wherever WebGPU is unavailable and served as the reference for testing the GPU engine.

\paragraph{Architecture.} The application is written in plain JavaScript and assembled from three sources into a single HTML file: the interface, the CPU engine and the GPU engine. Both engines expose the same interface, so the controls drive either. The Learn panels are written in \LaTeX{} notation and converted to MathML at build time, so equations render natively in current browsers without external libraries. The page needs no network access once loaded.

\section{Teaching with Condensate}
\label{sec:teaching}

\paragraph{The Learn panels.} Each model has a panel, aimed at graduate students, with four parts: \emph{the model}, giving the equations the application solves; \emph{the physics}, giving the key arguments and results; \emph{try this}, three or four activities; and \emph{read more}. A collapsible \emph{Foundations} section, common to all models, introduces the GPE, the hydrodynamic (Madelung) interpretation of density and phase, the Thomas--Fermi profile, the healing length and speed of sound with their values in the app, quantised vortices, and the numerical method.

\paragraph{Activities.} Each activity pairs a prediction that students can make with pen and paper with an experiment in the app. Examples include estimating the lattice spacing from Feynman's rule and checking it with the vortex markers; predicting the stripe period $\pi/k_1\approx1.8$ of the spin--orbit-coupled gas and the coupling $\Omega=2k_{so}^2$ at which the stripes vanish; predicting the miscibility threshold; computing the flat-top density of a quantum droplet; and deriving the Jeans threshold from the dispersion relation. Every quantitative prediction quoted in the panels was checked against the simulation. Some activities are deliberately open-ended. A ``Set this up'' button beside each activity configures the model, trap and parameters on the $128^2$ grid, so that the quoted numbers apply.

\paragraph{Links to the literature.} For convenience, the panels cite sections of the standard reference \emph{Quantum Fluids, Solitons, and Vortices}~\cite{bbp}, whose chapters on the GPE, waves, superfluidity, vortices, two-component and dipolar condensates, and droplets and supersolids map directly onto the models. Where the book does not cover a model, the panels cite review and primary literature instead: for spin--orbit coupling~\cite{lin,galitski}, ultralight dark matter~\cite{hui2017,hui2021}, polaritons~\cite{carusottociuti} and BKT physics~\cite{berezinskii,kosterlitz,hadzibabic,simula}. 


\paragraph{Limitations and Outlook.}  Condensate is a teaching tool, and its design makes compromises because of this. The models are fixed rather than user-defined and two-dimensional and the GPU engine runs in single precision. 
We would be interested in hearing your comments or criticisms, additions to the current models or further extensions that would prove useful to the community.

\section*{Availability}

Condensate is freely available from \url{https://abag.github.io/condensate/} and runs in most modern browsers; the GPU engine needs WebGPU, available in Chrome and Edge 113 or later, Safari 26 or later on macOS, iOS and iPadOS, and Firefox on Windows.

\section*{Acknowledgements}

The software the paper describes was developed with the assistance of Claude (Anthropic), a large language model, under the author's direction. We specified the physics, reviewed the code and derivations, directed validation tests and take full responsibility for the software and its description. We acknowledge fruitful conversations with Nick Parker and Ryan Doran, which improved the depth and functionality of the software.


\begin{thebibliography}{99}\small\setlength{\itemsep}{1pt}

\bibitem{bbp} C. F. Barenghi, T. Bland and N. G. Parker, \emph{Quantum Fluids, Solitons, and Vortices: A Primer with Worked Problems}, 2nd edn, Lecture Notes in Physics vol.\ 1050 (Springer, Cham, 2026), \href{https://doi.org/10.1007/978-3-032-20171-3}{doi:10.1007/978-3-032-20171-3}.
\bibitem{visualpde} B. J. Walker, A. K. Townsend, A. K. Chudasama and A. L. Krause, VisualPDE: rapid interactive simulations of partial differential equations, Bull.\ Math.\ Biol.\ \textbf{85}, 113 (2023), \href{https://doi.org/10.1007/s11538-023-01218-4}{doi:10.1007/s11538-023-01218-4}.
\bibitem{phet} C. E. Wieman, W. K. Adams and K. K. Perkins, PhET: simulations that enhance learning, Science \textbf{322}, 682 (2008).
\bibitem{explorables} D. Brockmann, Complexity Explorables, \url{https://www.complexity-explorables.org}.
\bibitem{kaboudian} A. Kaboudian, E. M. Cherry and F. H. Fenton, Large-scale interactive numerical experiments of chaos, solitons and fractals in real time via GPU in a web browser, Chaos Solitons Fractals \textbf{121}, 6 (2019).
\bibitem{quvis} A. Kohnle, M. Douglass, T. J. Edwards, A. D. Gillies, C. A. Hooley and B. D. Sinclair, Developing and evaluating animations for teaching quantum mechanics concepts, Eur.\ J.\ Phys.\ \textbf{31}, 1441 (2010).
\bibitem{gpelab} X. Antoine and R. Duboscq, GPELab, a Matlab toolbox to solve Gross--Pitaevskii equations I: computation of stationary solutions, Comput.\ Phys.\ Commun.\ \textbf{185}, 2969 (2014).
\bibitem{xmds} G. R. Dennis, J. J. Hope and M. T. Johnsson, XMDS2: fast, scalable simulation of coupled stochastic partial differential equations, Comput.\ Phys.\ Commun.\ \textbf{184}, 201 (2013).
\bibitem{gpue} J. Schloss and L. J. O'Riordan, GPUE: graphics processing unit Gross--Pitaevskii equation solver, J.\ Open Source Softw.\ \textbf{3}, 1037 (2018).
\bibitem{stagg} G. W. Stagg, webgl\_gpe: a WebGL implementation of a 2D Gross--Pitaevskii equation solver, \url{https://github.com/georgestagg/webgl_gpe}.
\bibitem{superhydro} M. M. Forbes et al., super\_hydro: superfluid hydrodynamics explorer, \url{https://github.com/mforbes/super_hydro}.
\bibitem{choi} S. Choi, S. A. Morgan and K. Burnett, Phenomenological damping in trapped atomic Bose--Einstein condensates, Phys.\ Rev.\ A \textbf{57}, 4057 (1998).
\bibitem{gaunt} A. L. Gaunt, T. F. Schmidutz, I. Gotlibovych, R. P. Smith and Z. Hadzibabic, Bose--Einstein condensation of atoms in a uniform potential, Phys.\ Rev.\ Lett.\ \textbf{110}, 200406 (2013).
\bibitem{ramanathan} A. Ramanathan et al., Superflow in a toroidal Bose--Einstein condensate: an atom circuit with a tunable weak link, Phys.\ Rev.\ Lett.\ \textbf{106}, 130401 (2011).
\bibitem{morsch} O. Morsch and M. Oberthaler, Dynamics of Bose--Einstein condensates in optical lattices, Rev.\ Mod.\ Phys.\ \textbf{78}, 179 (2006).
\bibitem{fischer} U. R. Fischer, Stability of quasi-two-dimensional Bose--Einstein condensates with dominant dipole--dipole interactions, Phys.\ Rev.\ A \textbf{73}, 031602(R) (2006).
\bibitem{ticknor} C. Ticknor, R. M. Wilson and J. L. Bohn, Anisotropic superfluidity in a dipolar Bose gas, Phys.\ Rev.\ Lett.\ \textbf{106}, 065301 (2011).
\bibitem{giovanazzi} S. Giovanazzi, A. G\"orlitz and T. Pfau, Tuning the dipolar interaction in quantum gases, Phys.\ Rev.\ Lett.\ \textbf{89}, 130401 (2002).
\bibitem{limapelster} A. R. P. Lima and A. Pelster, Quantum fluctuations in dipolar Bose gases, Phys.\ Rev.\ A \textbf{84}, 041604(R) (2011).
\bibitem{kadau} H. Kadau, M. Schmitt, M. Wenzel, C. Wink, T. Maier, I. Ferrier-Barbut and T. Pfau, Observing the Rosensweig instability of a quantum ferrofluid, Nature \textbf{530}, 194 (2016).
\bibitem{chomaz} L. Chomaz, I. Ferrier-Barbut, F. Ferlaino, B. Laburthe-Tolra, B. L. Lev and T. Pfau, Dipolar physics: a review of experiments with magnetic quantum gases, Rep.\ Prog.\ Phys.\ \textbf{86}, 026401 (2023).
\bibitem{aochui} P. Ao and S. T. Chui, Binary Bose--Einstein condensate mixtures in weakly and strongly segregated phases, Phys.\ Rev.\ A \textbf{58}, 4836 (1998).
\bibitem{petrov} D. S. Petrov, Quantum mechanical stabilization of a collapsing Bose--Bose mixture, Phys.\ Rev.\ Lett.\ \textbf{115}, 155302 (2015).
\bibitem{cabrera} C. R. Cabrera, L. Tanzi, J. Sanz, B. Naylor, P. Thomas, P. Cheiney and L. Tarruell, Quantum liquid droplets in a mixture of Bose--Einstein condensates, Science \textbf{359}, 301 (2018).
\bibitem{semeghini} G. Semeghini et al., Self-bound quantum droplets of atomic mixtures in free space, Phys.\ Rev.\ Lett.\ \textbf{120}, 235301 (2018).
\bibitem{petrovastrakharchik} D. S. Petrov and G. E. Astrakharchik, Ultradilute low-dimensional liquids, Phys.\ Rev.\ Lett.\ \textbf{117}, 100401 (2016).
\bibitem{lin} Y.-J. Lin, K. Jim\'enez-Garc\'ia and I. B. Spielman, Spin--orbit-coupled Bose--Einstein condensates, Nature \textbf{471}, 83 (2011).
\bibitem{hozhang} T.-L. Ho and S. Zhang, Bose--Einstein condensates with spin--orbit interaction, Phys.\ Rev.\ Lett.\ \textbf{107}, 150403 (2011).
\bibitem{lipitaevskiistringari} Y. Li, L. P. Pitaevskii and S. Stringari, Quantum tricriticality and phase transitions in spin--orbit coupled Bose--Einstein condensates, Phys.\ Rev.\ Lett.\ \textbf{108}, 225301 (2012).
\bibitem{galitski} V. Galitski and I. B. Spielman, Spin--orbit coupling in quantum gases, Nature \textbf{494}, 49 (2013).
\bibitem{sonstephanov} D. T. Son and M. A. Stephanov, Domain walls of relative phase in two-component Bose--Einstein condensates, Phys.\ Rev.\ A \textbf{65}, 063621 (2002).
\bibitem{hui2017} L. Hui, J. P. Ostriker, S. Tremaine and E. Witten, Ultralight scalars as cosmological dark matter, Phys.\ Rev.\ D \textbf{95}, 043541 (2017).
\bibitem{hui2021} L. Hui, Wave dark matter, Annu.\ Rev.\ Astron.\ Astrophys.\ \textbf{59}, 247 (2021).
\bibitem{schive} H.-Y. Schive, T. Chiueh and T. Broadhurst, Cosmic structure as the quantum interference of a coherent dark wave, Nature Phys.\ \textbf{10}, 496 (2014).
\bibitem{woucar} M. Wouters and I. Carusotto, Excitations in a nonequilibrium Bose--Einstein condensate of exciton polaritons, Phys.\ Rev.\ Lett.\ \textbf{99}, 140402 (2007).
\bibitem{carusottociuti} I. Carusotto and C. Ciuti, Quantum fluids of light, Rev.\ Mod.\ Phys.\ \textbf{85}, 299 (2013).
\bibitem{wousavona} M. Wouters and V. Savona, Stochastic classical field model for polariton condensates, Phys.\ Rev.\ B \textbf{79}, 165302 (2009).
\bibitem{gardinerdavis} C. W. Gardiner and M. J. Davis, The stochastic Gross--Pitaevskii equation: II, J.\ Phys.\ B \textbf{36}, 4731 (2003).
\bibitem{bradley} A. S. Bradley, C. W. Gardiner and M. J. Davis, Bose--Einstein condensation from a rotating thermal cloud: vortex nucleation and lattice formation, Phys.\ Rev.\ A \textbf{77}, 033616 (2008).
\bibitem{blakie} P. B. Blakie, A. S. Bradley, M. J. Davis, R. J. Ballagh and C. W. Gardiner, Dynamics and statistical mechanics of ultra-cold Bose gases using c-field techniques, Adv.\ Phys.\ \textbf{57}, 363 (2008).
\bibitem{bao} W. Bao, D. Jaksch and P. A. Markowich, Numerical solution of the Gross--Pitaevskii equation for Bose--Einstein condensation, J.\ Comput.\ Phys.\ \textbf{187}, 318 (2003).
\bibitem{strang} G. Strang, On the construction and comparison of difference schemes, SIAM J.\ Numer.\ Anal.\ \textbf{5}, 506 (1968).
\bibitem{baoduzhang} W. Bao, Q. Du and Y. Zhang, Dynamics of rotating Bose--Einstein condensates and its efficient and accurate numerical computation, SIAM J.\ Appl.\ Math.\ \textbf{66}, 758 (2006).
\bibitem{chiofalo} M. L. Chiofalo, S. Succi and M. P. Tosi, Ground state of trapped interacting Bose--Einstein condensates by an explicit imaginary-time algorithm, Phys.\ Rev.\ E \textbf{62}, 7438 (2000).
\bibitem{berezinskii} V. L. Berezinskii, Destruction of long-range order in one-dimensional and two-dimensional systems possessing a continuous symmetry group. II. Quantum systems, Sov.\ Phys.\ JETP \textbf{34}, 610 (1972).
\bibitem{kosterlitz} J. M. Kosterlitz and D. J. Thouless, Ordering, metastability and phase transitions in two-dimensional systems, J.\ Phys.\ C \textbf{6}, 1181 (1973).
\bibitem{hadzibabic} Z. Hadzibabic, P. Kr\"uger, M. Cheneau, B. Battelier and J. Dalibard, Berezinskii--Kosterlitz--Thouless crossover in a trapped atomic gas, Nature \textbf{441}, 1118 (2006).
\bibitem{simula} T. P. Simula and P. B. Blakie, Thermal activation of vortex--antivortex pairs in quasi-two-dimensional Bose--Einstein condensates, Phys.\ Rev.\ Lett.\ \textbf{96}, 020404 (2006).

\end{thebibliography}
\end{document}